 \documentclass[final,5p,twocolumn]{elsarticle_arXiv}
\usepackage{graphics}
\usepackage{amssymb}
\usepackage{amsmath}
\usepackage{latexsym}
\usepackage{supertabular}
\usepackage{slashed}

\newcommand{\nn}{\nonumber}
\newcommand{\sla}{\slashed}

\journal{arXiv:hep-ph}

\begin{document}
\renewcommand{\include}[1]{}
\renewcommand\documentclass[2][]{}
\begin{frontmatter}


\preprint{KEKTH-1546}

\title{On the W+4 jets background to the top quark asymmetry at the Tevatron}

\author[add1]{Kaoru Hagiwara}
\ead{kaoru.hagiwara@kek.jp}
\author[add2]{Junichi Kanzaki}
\ead{junichi.Kanzaki@cern.ch}
\author[add1]{Yoshitaro Takaesu}
\ead{takaesu@post.kek.jp}

\address[add1]{KEK Theory Center and Sokendai, Tsukuba, 305-0801, Japan}
\address[add2]{KEK, Tsukuba, 305-0801, Japan}

\begin{abstract}
We investigate the possibility that the
 $W$+4\,jets background is a source of the anomalously large
 top-antitop ($t\bar{t}$) charge asymmetry observed at the Tevatron. We
 simulate the $t\bar{t}$
reconstruction of the signal and background events at the
 matrix-element level and find that the reconstructed $t\bar{t}$
candidates from the $W$+4\,jets background could
give large forward-backward asymmetry.
We suggest serious re-evaluation of the $W$+4\,jets background for the
 $t\bar{t}$ candidate events by studying the distributions of the reconstructed
 $t\bar{t}$ systems in their rapidity
difference, the transverse momentum of the
 top
 quark, and that
of the $t\bar{t}$ system, which can differ significantly from those of the QCD
 $t\bar{t}$ production events especially at high $t\bar{t}$ invariant mass.
\end{abstract}

%\begin{keyword}


%\end{keyword}

\end{frontmatter}

% \linenumbers

\documentclass[final,5p,twocolumn]{elsarticle}

\section{Introduction}
% Importance of the research theme
% Conventional research results
% Unresolved points in conventional research
% Purpose
% Method
% Results
% Conclusion
% Order of configuration of paper
\label{intro}
Since CDF\cite{Aaltonen:2011kc} and D0\cite{Abazov:2011rq} experiments
reported the significant deviation
from the next-to-leading order (NLO)
prediction\cite{Kuhn:1998kw,Bowen:2005ap,Dittmaier:2007wz,Almeida:2008ug} in
the forward-backward asymmetry of top-antitop ($t\bar{t}$) productions in $p\bar{p}$ collisions, the
origin of the discrepancy has been one of the serious issues in high-energy physics. Contributions from the NNLL QCD\cite{Kidonakis:2011zn} and the QED
higher order corrections\cite{Hollik:2011ps}  have been
discussed, but they are not
sufficient to explain the anomaly. Although both experiments seem to be
confident that the
large asymmetry is originated in the $t\bar{t}$ signal events, there may be room for
the background events to be responsible for the large positive asymmetry
in view of the complexity of the
experimental analysis and our imperfect understanding of multi-jet
events at hadron colliders.

In this short report we investigate the possibility that the $W$+4 jets background, the main
background for the $t\bar{t}$ production process, is a source of the
disagreement in the top forward-backward asymmetry. We simulate the experimental
event-selection cuts for the $t\bar{t}$-candidate events using the event
samples generated with exact
leading-order matrix elements. It is then found that the reconstructed $t\bar{t}$
candidates from $W$+4 jets background can give rise to the large
forward-backward asymmetry, especially
at high $t\bar{t}$ invariant mass. We present kinematic distributions
at the parton level because the smearing due to subsequent parton
showering is not fully understood for multi-jet processes at hadron
colliders, as hinted by the discrepancy in the di-jet invariant
mass distribution in $W$+2 jets events studied at the
Tevatron\cite{Aaltonen:2011mk,Abazov:2011af}. The following results
should hence be regarded as hints of what could be observed in real
experiments after smearing due to QCD shower development and
hadronization, detector resolutions, and event reconstruction complexities.
\bibliographystyle{bib/model1a-num-names}
\bibliography{bib/reference}
\end{document}